# Technologies for promoting social participation in later life

**Authors**
Marcos Baez, University of Trento and Tomsk Polytechnic University
Radoslaw Nielek, Polish-Japanese Academy of Information Technology
Fabio Casati, University of Trento and Tomsk Polytechnic University
Adam Wierzbicki, Polish-Japanese Academy of Information Technology

**Abstract**
Social participation is known to bring great benefits to the health and well-being of people as they age. From being in contact with others to engaging in group activities, keeping socially active can help slow down the effects of age-related declines, reduce risks of loneliness and social isolation and even mortality in old age. There are unfortunately a variety of barriers that make it difficult for older adults to engage in social activities in a regular basis. In this chapter, we give an overview of the challenges to social participation and discuss how technology can help overcome these barriers and promote participation in social activities. We examine two particular research threads and designs, exploring ways in which technology can support co-located and virtual participation: i) an application that motivates the virtual participation in group training programs, and ii) a location-based game that supports co-located intergenerational ICT training classes. We discuss the effectiveness and limitations of various design choices in the two use cases and outline the lessons learned.[1]

**Highlights**

- Technology can help overcome social participation challenges faced by older adults and facilitate social inclusion via virtual and co-located activities.

- Technology design and evaluation should consider the diversity of the older adult population, not only in terms of abilities but individual and cultural differences that can shape social participation.

- Enabling social participation does not guarantee actual participation, so it is important that sociotechnical systems have an active role in engaging and motivating social interactions.

---

[1] The present is an author copy, published under the self-archiving policy by Springer. Cite this paper as:
Baez M., Nielek R., Casati F., Wierzbicki A. (2019) Technologies for Promoting Social Participation in Later Life. In: Neves B., Vetere F. (eds) Ageing and Digital Technology. Springer, Singapore

## 1.1 Introduction

In this chapter we study the opportunities and means through which technology can improve our general wellbeing by enabling us to engage in social activities that meet our needs, interests and allow us to stay active as we age. Social participation, from being in contact with others to engaging in sharing activities and contributing to society, is indeed a fundamental modifiable determinant (Levasseur, Richard, Gauvin, & Raymond, 2010) that has been associated not only with happiness and wellbeing (Graney, 1975) but also with health, morbidity and mortality in later age (Berkman, Glass, Brissette, & Seeman, 2000; Levasseur, Desrosiers, & Tribble, 2008; Tilvis, Laitala, Routasalo, & Pitkälä, 2011). Scant social participation puts older adults at risk of loneliness and social isolation (Pinquart & Sorensen, 2001), and their devastating effects on physical and mental health, e.g., increased mortality rates, elevated blood pressure, dementia, depression, and cognitive decline (Bower, 1997; Fratiglioni, Wang, Ericsson, Maytan, & Winblad, 2000; Heikkinen & Kauppinen, 2004).

Despite the benefits of social participation, worldwide trends in loneliness and social isolation[2] show a situation that is widespread and increasing, especially in developed countries. Surveys from USA (Edmondson, 2010), China (Yang & Victor, 2008) and Europe (Yang & Victor, 2011) report social isolation and loneliness in older adults on these different geographical and cultural regions with measures varying depending on the country and the scale used, with more marked presence in the oldest old (Yang & Victor, 2011). For instance, the results of a nationwide survey in Finland (2002) with a sample of 3858 community-dwelling older adults (75+) (Tilvis et al., 2012), show that 46% of older adults are socially isolated and 37% experience loneliness. Loneliness has been increasing (from 20% to 35%) in the US in just a decade (2000 - 2010) (Edmondson, 2010) for people 45+, and similarly (from 15% to 29%) in China in 8 years for older adults 60+ (national surveys of older people done in 1992 and 2000) (Yang & Victor, 2008). This trend gives us a hint to the dimension and extent of the barriers and challenges to social participation.

Given this context, a growing body of interdisciplinary research has been focusing on how to facilitate social participation as people age, and specifically of how technology can support people in remaining socially active even in the wake of physical, cognitive, and mobility challenges. In order to better understand the role of technology in this space, it is important to consider the two type of interpersonal interactions (Tong & Walther, 2011): *virtual* (i.e., communication over a distance) or *co-located* (i.e., face-to-face) . This distinction is essential as each is built on different assumption in terms of abilities and opportunities of older adults.

---

[2] Notice that while loneliness and social isolation are often used interchangeably, they refer to different yet interlated concepts: loneliness is a *subjective* measure of the "unpleasant" response to the lack of social relationships (de Jong Gierveld, van Tilburg, & Dykstra, 2006) and social isolation an *objective* measure referring to the lack (absence or low number) of social relationships (de Jong Gierveld, van Tilburg, & Dykstra, 2006).

In this chapter, we first discuss barriers to social participation and then present results from two research threads addressing the problem of how technology can promote social participation in virtual and co-located environments. In the first use case, we review a tablet application that motivates virtual participation in social group training programs, while in the second, a location-based game that supports co-located intergenerational ICT training classes. We conclude by providing some lessons learned and outlining opportunities for further research.

## 1.2 Challenges to social participation

Engaging in social activities is known to bring great benefits to the well-being of people as they age (Graney, 1975). This is true for a wide range of activities, most notably for volunteering (Graney, 1975; Musick & Wilson, 2003), exercising (Spirduso & Cronin, 2001; Stuart, Chard, Benvenuti, & Steinwachs, 2008), leisure activities (Menec & Chipperfield, 1997; Ragheb & Griffith, 1982), and visiting friends and family (Graney, 1975; Montross et al., 2006).

There are, however, a variety of barriers that prevent older adults from engaging in social activities, bringing undesired effects on their health and well-being (Berkman et al., 2000; R. S. Tilvis et al., 2011). Unfortunately, overcoming these barriers is usually beyond the affected person's control (Wenger, Davies, Shahtahmasebi, & Scott, 1996), requiring the support of special intervention programs and services, and lately, opening up opportunities for technology support.

Interventions programs to enable and promote social participation by older adults are commonly challenged by the following barriers:

- *Mobility constraints*. The ability to get out and move around one's environment is fundamental to active aging (Webber, Porter, & Menec, 2010). Age-related diseases and functional problems, however, pose mobility constraints that significantly affect older adults' social participation and engagement (Rosso, Taylor, Tabb, & Michael, 2013). Webber et al. (2010) go beyond functional abilities to define mobility in older adults as a complex concept described by five categories of determinants (cognitive, psychosocial, physical, environmental and financial). Issues in these determinants and related factors reduce the ability of older adults to take an active role in social participation.

- *Lack of companions*. The social network of a person changes across the lifespan, getting smaller as we age, both in terms of personal and friendship networks (Wrzus, Hänel, Wagner, & Neyer, 2013), and geographical proximity (Ajrouch, Blandon, & Antonucci, 2005). These changes, along with life events, such as retirement or bereavement, may also limit the social participation of older adults for lack of available companions (Havens,

Hall, Sylvestre, & Jivan, 2004). Additionally, these factors put older adults at the risk of loneliness and social isolation (Pinquart & Sorensen, 2001).

- *Lack of motivation*. A common barrier to sustaining active participation to social activities is lack of motivation. For example, we know this is important for engaging in volunteering (Wilson & Musick, 1999) and physical activities (de Groot & Fagerström, 2011) — activities that are known to be beneficial and to provide opportunities for social contact. Indeed, motivation is often associated with attrition rates and adherence, metrics used to measure the effectiveness of intervention programs in general, including those aiming at increasing the well-being of older adults (Cattan, White, Bond, & Learmouth, 2005).

- *Lack of opportunities*. Another limitation to engaging in social activities is simply lack of opportunities. Older adults might be living in communities that are not "aging-friendly", lacking the necessary support that would allow them to engage in activities of their interest and meet their social needs, through appropriate products or services (Scharlach, 2012).

- *Lack of IT-skills*. The use of technology opens up opportunities for older adults to stay in touch with family and friends, especially for those with more limitations to participate in social activities (Barnard, Bradley, Hodgson, & Lloyd, 2013). However, due to the distinct abilities of the older adult population (Charness & Bosman, 1990), use and adoption of technology is a recognized challenge amongst this group (Barnard et al., 2013; Peek et al., 2016).

Technology has the potential to overcome the above barriers and provide support to those more challenged, creating an auspicious platform for social participation (DiMaggio, Hargittai, Neuman, & Robinson, 2001; Haythornthwaite, 2005). While there are many ways in which we can characterize technological support, in this chapter we borrow the classification from computer-mediated communication research (Tong & Walther, 2011) to consider how social participation is mediated by technology:

- *Technology for virtual participation,* providing support for geographically distant participation. The level of support is usually characterized by the richness of the medium (Daft & Lengel, 1986), to denote the degree at which the medium can carry non-verbal cues. From email, to video conferencing to virtual reality, technology has been developing to provide different levels of social presence and media richness.

- *Technology for co-located participation,* providing support for co-located activities. This is growing area of research focusing on how technology can augment the experiences of individuals and groups. Whether exercising with a friend using the Nintendo Wii game console, reminiscing on pictures with the family, technology for co-located activities are showing potential benefits for older adults (Chao, Scherer, & Montgomery, 2015; Lazar, Thompson, & Demiris, 2014).

In what follows we build on the characterization of barriers and technology support to analyze two research threads we undertook in IT-support for social participation:
i) technology for virtual participation in group exercising, and ii) technology for co-located participation in ICT learning.

## 1.3 Virtual participation to group exercising

Engaging in physical activity can bring multiple benefits to the health and well- being of older adults (Spirduso & Cronin, 2001). It reduces risk of falls (Thibaud et al., 2012), slows progression of degenerative diseases (Stuart et al., 2008), and even improves cognitive performance and mood (Landi et al., 2010). However, engaging in regular physical activity can be challenging for some older adults for the same reasons they find it difficult to engage in social activities. Thus, and in spite of the growing evidence of the benefits of physical activity, as well as the adverse effects of sedentary behavior (Wilmot et al., 2012), physical inactivity is still prevalent in older adults (Harvey, Chastin, & Skelton, 2013).

Technology for fitness training, ranging from DVDs (Wójcicki et al., 2014) to tablet applications (Silveira et al., 2013), and increasingly, gaming technology (Carmichael, Rice, & MacMillan, 2010), have been used to facilitate home-based training for older adults. However, most solutions for older adults downplay the importance of the social context as a motivating factor in physical training (Far, Nikitina, Baez, Taran, & Casati, 2016) and ignore the opportunities of shared activities as a platform for social interactions. This represents a limitation in current systems, as previous studies suggest not only that older adults prefer exercising with others rather than individually (de Groot & Fagerström, 2011), but also that a social context can lead to higher levels of participation (Silveira et al., 2013).

In this use case we describe a home-based training application, namely *Gymcentral*, that supports virtual participation in group-exercising, enabling older adults, who for various reasons are not able to join in person group training, to keep physically and socially active from home.

### 1.3.1 Gymcentral

Gymcentral is a platform and a tablet-based fitness environment designed to keep independent-living older adults physically and socially active. It does so by providing trainees with a virtual environment that is both personal, i.e., the training program and feedback are personalized, and social, i.e., members can interact and participate to group exercise sessions even if they have different physical abilities. The application is based on years of research on home-based training (Báez, Ibarra, Far, Ferron, & Casati, 2016; Far, Ibarra, Baez, & Casati, 2014; Far, Silveira, Casati, & Baez, 2012; Silveira et al., 2013).

The Gymcentral platform is organized in two main applications that serve the needs of both trainees and the coach. The Coach App is a web-based system that allows the training expert to define, monitor and adapt the training programs remotely, from a computer or a tablet. The Trainee App is a tablet application (for iPad and Android) that allows trainees to follow the training programs from the comfort of house, following video instructions set by the Coach in a virtual social classroom on what and how exercises should be performed. Together, these applications can support a typical workflow as illustrated in **Error! Reference source not found.** and described next:

- The Coach defines a training program, which includes exercise intensity levels, instruction videos and a training schedule.
- The Coach then assesses the aptitude of each individual trainee, assigns an intensity level profile and further tailors the program in case of special needs.
- Trainees follow the training program from home using the tablet application. They received a tailored exercise program that fits their abilities.
- Users can participate in virtual group exercise sessions despite their different abilities. Each can see the other trainees in the virtual gym and invite those not present to join the training session.
- Trainees self-report on their performance (based on questionnaires defined by the Coach) or automatically via application logs and sensors.
- The Coach can see the progress of the trainees, give personalised feedback, and decide on whether to increase the intensity of the exercises for each individual trainee.
- The Coach can intervene at any point and tune the individual programs, e.g., in case a trainee is experiencing pain, and keep track of any particular event in an online diary.
- Trainees can contact the Coach for support, and interact with each other via private and public messages. In the same way, the Coach can participate in the public discussions to build a sense of community and motivate the trainees.
- Trainees can keep track of their own progress via progress metaphors.

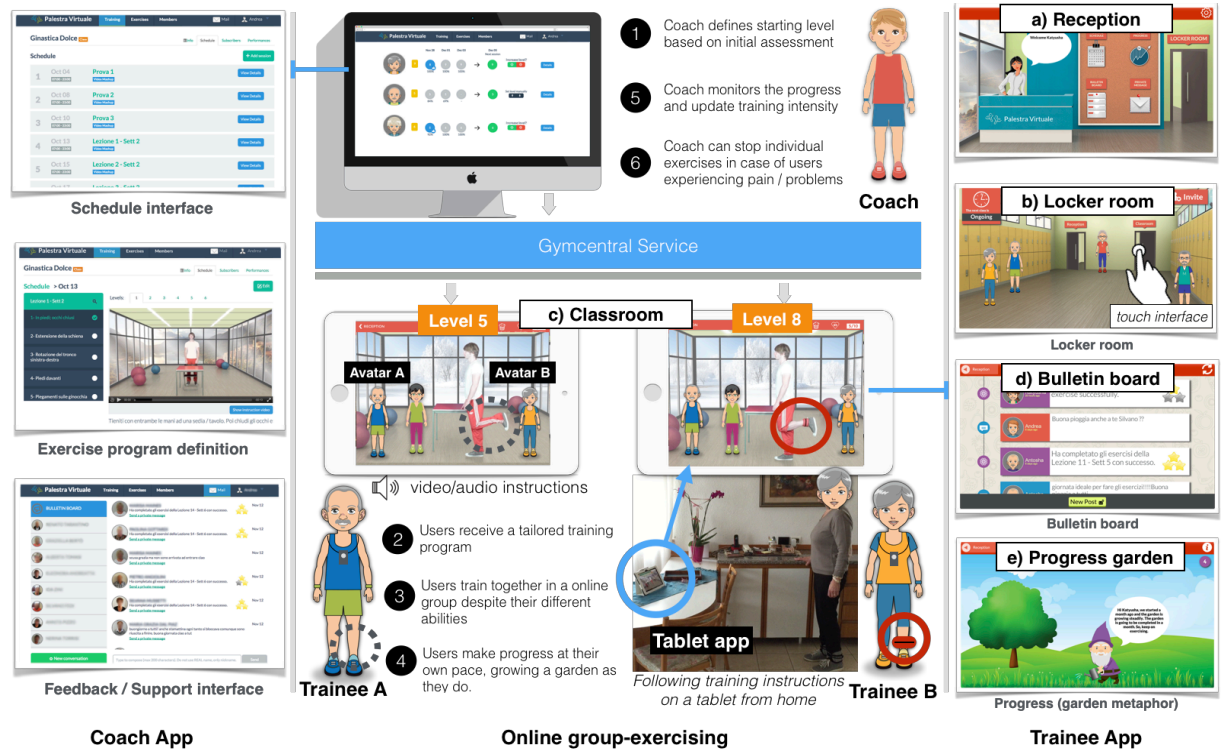

**Figure 0.1: Overview of the Gymcentral service.** The figure highlights the main areas of the Trainee app: a) *reception*, the entry point to all the services of the gym, b) *locker room*, the space where users can meet before the training and have contextual interactions, c) *classroom*, the space where users can join together a training session and be aware of the coach as well as other trainees, d) *messaging*, feature that allows public and private messages, e) *progress*, feature that allows visualising own' progress using the metaphor of a growing garden.

### 1.3.2 Design rationale

In developing an effective support for group-exercising, our design goal was to enable and motivate older adults to exercise from home with the virtual company of training companions (e.g., friends or other gymcentral members). In this section we explain our approach to overcoming the participation barriers and the design choices we took in the development process. A discussion on the impact of these choices is presented in section 1.3.3.

*Relying on known metaphors to address those with low ICT skills.*
Technologies should be designed to fit the needs and abilities of the target population. In this case, it required us to consider the possibility of users with low levels of computer literacy and potentially age-related declines (Charness & Bosman, 1990). While there are several design guidelines for older adults (Fisk, Czaja, Rogers, Charness, & Sharit, 2009; Kurniawan & Zaphiris, 2005), the one that percolated to the core of our design was the use of known or relatable metaphors. The design of Gymcentral relies on the metaphor of a virtual gym, providing similar spaces and services found in a real gym (**Error! Reference source not**

**found.**): a reception, the entry point of the Gym, where the user has access to all the services, locker room, a space where trainees usually meet each other, invite those missing, and get ready for the training classes, and a classroom, a space where users have access to the exercise instructions and train together. Navigation between these spaces is done by interacting with known objects, such as tapping on an open door to enter the classroom – if the class is open, otherwise a closed door is shown indicating that no training sessions are ongoing – or tapping indicative icons on the reception board to check their message or training agenda. In organizing the design around these virtual spaces, and providing navigational and visual cues, we aimed at facilitating the navigation and usage of Gymcentral by older adults with (almost) no ICT skills.

*Enabling virtual participation from home.*
Older adults with mobility constraints are more limited in their opportunities to engage in group exercising. The idea of the virtual group-exercising is to overcome this barrier and enable older adults to enjoy the benefit of exercising in a social context from their home. To enable virtual social participation, the design relies on virtual environments, which have been shown to increase the sense of presence, or psychological immersion (Grinberg, Careaga, Mehl, & O'Connor, 2014). In addition, social presence, along with user embodiments (avatars), help to reduce physical barriers and get users more engaged in the activities while preserving their privacy (Siriaraya, Ang, & Bobrowicz, 2014). Avatars however do not mimic the actual trainees' movement during the exercises but follow predefined animated movements. This was both a practical constraint (i.e., to keep the technological requirement to a minimum) and a design constraint (i.e., to keep the specifics of the exercise performed hidden from others) to avoid the negative effects of face-to-face group exercising in heterogenous groups (e.g., limited effectiveness and lack of motivation) (de Groot & Fagerström, 2011).

*Creating opportunities for virtual social interaction.*
Engaging in activities with others can help stimulate social interactions (Leonardi, Mennecozzi, Not, Pianesi, & Zancanaro, 2008). This is particularly beneficial for older adults with limited opportunities to interact — in most cases for the same reasons they need home training. Training together could then potentially help older adults to stay physically and socially active. We build on this opportunity by providing three different channels: the bulletin board, private messages, and contextual interactions in virtual spaces (see **Error! Reference source not found.**). The bulletin board is a community feature where trainees can exchange public messages. Performance and exercise achievements of the trainees are also automatically published on the bulletin board. Similar to the bulletin board, private messaging enables trainees to post and receive messages from the coach and other trainees, although only as one-to-one communication channel. Ephemeral interactions in the locker room enable trainees to engage in quick and contextual interactions, e.g., users can see each other (as avatars) in the locker room and interact by means of predefined messages (e.g. "Hi, let's go to the classroom"). In providing different social interaction channels, we aimed not only at offering a choice of communication but also at observing emerging social interaction patterns.

*Stimulating participation and program adherence.*
Self-efficacy (i.e., perceived capability and confidence), a strong predictor of adherence to physical exercises, is less exhibited in older adults compared to other age groups (Phillips, Schneider, & Mercer, 2004). Studies have shown that the use of persuasive features (especially social persuasion strategies) increases the adherence to training programs (Silveira, van de Langenberg, et al., 2013). Gymcentral incorporates individual and social persuasion strategies derived from previous work on persuasion (among others, (Fogg, 2002; Oinas-Kukkonen & Harjumaa, 2008)), implemented in the application as described below:

- *Individual persuasion strategies*, such as Self-monitoring, by giving trainees an awareness about their current progress, visualized using the garden metaphor; and positive & negative reinforcement, by prompting positive or negative comments about the exercising behavior of the trainee to raise awareness after a training session.

- *Social persuasion strategies*, such as social learning, by allowing trainees to compare their performance with others, social support, by enabling trainees to create a community of people supporting each other, social facilitation, by providing social spaces like the locker room and the classroom that allow for social awareness, and normative influence, by allowing users to send and receive invitations to exercise together, thus acting as a peer pressure mechanism. This strategy aims to address motivational issues affecting engagement in group exercising activities.

### 1.3.3 Studies and Findings

The feasibility and effectiveness of Gymcentral, as a tool to enable and motivate the participation older adults in home-based physical interventions, has been the subject of several randomised control trials in Italy, Netherlands, and Russia (Baez et al., 2017; Geraedts et al., 2017; Nikitina, Didino, Baez, & Casati, 2018). The results consistently show the feasibility of the tool, and provide further insights into the effects of the various design choices and their limitations in promoting social participation. In what follows we summarise the findings, focusing on the results of the study in Trento (Baez et al., 2017) for simplicity, although the overall results have been consistent across studies. This was a randomised pilot trial with a total of 37 older adults aged between 65 and 87 years old, who followed a personalized fall prevention exercise program for a period of eight weeks. Participants were randomly assigned to an *intervention* condition with access to the full features of the Trainee App, and a *control* condition with access to a simplified version limited to individual training. All participants were supported remotely by a professional coach.

*Usability and technology acceptance.*
We studied the technology acceptance and perceived usability of Gymcentral, exploring how it evolved over time (pre and post study) and compared to a simpler version limited to individual training. Not surprisingly, the usability was lower for Gymcentral at the beginning of the study, reflecting participants initial difficulties to deal with a more complex user interface. However, by the end of the intervention program, perceived usability had increased significantly, approaching the top end of the scale and the performance of the simpler application. Overall, while Internet connection was an intermittent issue, the usability and technology acceptance of both applications (group-exercising and individual-training versions) generally improved. For Gymcentral, these results mean that users could handle the extra complexity and learn to use this type of tool (Baez et al., 2016).

*Feasibility of the virtual participation.*
We investigated if, given the possibility, trainees would choose to virtually train together as opposed to training alone. Thus, in the studies we gave trainees the option to participate in the training session at any time, joining other users in group trainings sessions or exercising alone. We set as a control condition a group of participants using the individual training application, without mutual awareness, as to capture meetings by chance. The results showed a significant difference in co-presence (i.e., training sessions where the participant exercised with the company of at least one other trainee, see Far et al., 2015) in the intervention group, where social presence was in place in the form of virtual avatars, compared to the meetings by chance in the control group (Far et al., 2015). In addition, the success rate of the feature to invite others to join was encouraging, adding to the evidence on the preference of group-exercising, and motivating further study into the effects of normative influence in co-presence. These observations are reinforced by the user feedback on the value of group-exercising (Baez et al., 2016).

*Nature of virtual interactions.*
We studied if and how trainees made use of virtual interactions channels during the trials (Baez et al., 2016). Our observations show that the bulletin board was used mainly to promote community building, where the participants had an active role supporting each other. The distinctive use of the private messages was for clarifying questions regarding the training and receiving support from the Coach, as well as for personal support messages among trainees. These results highlighted the need for having both types of channels, since they serve different purposes.

We should also note that compared to other technology-based interventions, where social features (e.g., forums or social networks) were rarely used (Aalbers, Baars, & Rikkert, 2011), in our study the social features were largely used by the participants. However, we were not successful at motivating contextual messages in virtual spaces, and this is evident in the low number of contextual interactions and the low perceived usefulness of the feature by the participants. This points to the need for more effective environments for motivating real-time social interactions.

At the end of the intervention, control and intervention groups showed increased subjective wellbeing and reduced loneliness levels (see Baez et al., 2017 for instruments and measures). Both groups observed these benefits (related to regular physical activity) despite the presence of social interaction features only in the Intervention group application. Further analysis showed a moderate negative correlation between loneliness levels and the number of private messages exchanged by the participants, suggesting an association between virtual interactions and improvement in social wellbeing. The improvement in the control group was unexpected but can be attributed to the weekly calls by the Coach to provide support.

*Persuasion strategies and adherence to a training program.*
We studied the effect of the persuasion strategies, and in particular of social persuasion strategies, on the adherence of trainees to a training program (Far et al., 2015). The results indicate that participants training with the support of persuasion strategies feature a significantly higher participation in training sessions compared to participants without such support. Furthermore, we have observed that trainees have not only complied with the minimum attendance requirement by the Coach but attended even more training sessions. These results are encouraging as they suggest that the effects of the application are not limited to compliance but promote real engagement.

## 1.4 Co-located participation mediated by ICT

ICT technologies are frequently used to facilitate remote interaction of older adults, e.g. discussion about travelling (Balcerzak & Nielek, 2017) or programming (Kowalik & Nielek, 2016) but are not limited to it. Nielek, Lutostanska, Kopec, & Wierzbicki (2017) have studied the possibility to contribute to Wikipedia by older adults but one of the most promising areas is to enriching co-located participation. According to the study conducted by Gajadhar, Nap, de Kort, and IJsselsteijn (2010) face-to-face interactions are seen by older adults as high-quality social activities.

Researchers developing such technologies dedicated to older adults need to address problems such as low technology acceptance and lack of ICT skills, but there are at least a few examples showing that it is doable. Ceriani, Bottoni, Ventura, and Talamo (2014) developed a platform composed of dedicated application running on interactive table to support sharing life experiences and participatory story telling by collaborative production of video content. In a study conducted by Pedell, Beh, Mozuna, and Duong (2013) design and testing of multi-players games for co-located playing were preceded by a five-week long training program.

ICT are particularly effective for fostering co-located intergenerational participation (Chua, Jung, Lwin, & Theng, 2013). A cooperative two-players silhouette game developed by Rice and colleagues (Rice et al., 2013; Rice, Yau, Ong, Wan, & Ng, 2012) helped to reveal that although pairs composed of two older adults communicate, more mixed pairs do more physical cooperation. General positive effects of intergenerational co-located playing were observed not only for specially designed games, but also for standard off-the-shelf games like "Wi Sports" (Theng, Chua, & Pham, 2012).

Next to physical exercising and social participation, learning in later life has also overwhelmingly positive health and emotional effects (Aldridge & Lavender, 2000; Dench & Regan, 2000). It can be instrumental in enabling older adults to engage in volunteering, civic activities, and take social actions (Githens, 2007) by providing them with the required skills. In an online setting, it can even beneficial those with health or mobility problems (Chaffin & Harlow, 2005).

This use case is based on a location-based game that can be thought of as a tool for teaching older people mobile device technology (use of tablets) using the learning-by-doing approach but, at the same time, fostering co-located intergenerational social activity.

### 1.4.1 Location based game

The location-based-game research case described in this section is a part of the Living Laboratory project (LivingLab; detailed description can be found in (Kopec et al., 2017)) initialized, developed and implemented at the Polish-Japanese Academy of Information Technology (PJAIT) located in Warsaw, Poland. The LivingLab is run in cooperation with the Municipality of Warsaw. LivingLab goals address vital problems of social informatics, especially research and development of solutions for active aging and healthy living, game application for better lifestyle and well-being, positive gaming, stress management and technologies enhancing social well-being. Currently, the LivingLab has over 200 older participants, most of whom are seniors who completed a basic computer course provided by the City of Warsaw.

The location-based game aimed at tackling several crucial topics related to the aging. We have explored issues of social inclusion in later life as well as education of older adults in the field of mobile technologies and their motivation to learn. At the same time, we have organized physical activity for older adults. The tool that allowed us to combine all these treatments was a location-based game "Stroll Around Yesterday" that combined historical knowledge with the use of tablets and interaction in mixed-age teams of two players (a senior and a PJAIT student of computer science). We also chose this form of activity to help students understand the requirements that should be taken into account in the process of creating software applications for senior citizens.

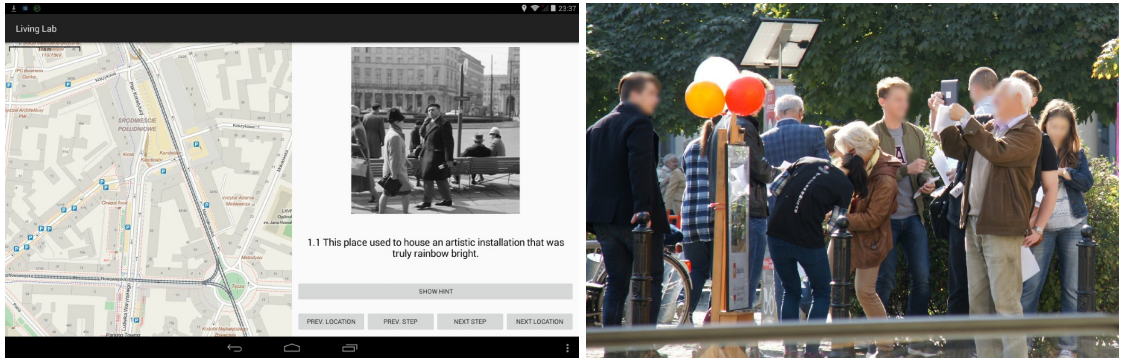

**Figure 0.2: Overview of the game "Stroll around yesterday".** The left figure shows a screen shoot from the game "Stroll around yesterday". The screen of application is divided vertically by two parts. On the left side a current map of Warsaw is displayed and accompanied, on the right side, by an old picture of Warsaw and navigation buttons. The right figure presents a picture taken during one of the gameplay sessions.

The game setup was inspired by the study of related work and literature supported by a set of best practices conveyed by external consultants experienced in location-based game design and senior outdoor activities (e.g., city tour guides). The game "Stroll Around Yesterday" requires the use of tablets and interaction in mixed-age teams of two players: a senior and a junior. The study concept was to stimulate interaction and cooperation between the team partners: on the one hand, the older participants were using the device and mobile apps with an indirect assistance of the younger tech-savvy team member, on the other hand, the older adult should be more familiar with the historical and cultural context of the game (location descriptions and hints based on the literature and photos from the past).

### 1.4.2   Design rationale

The design of the location-based game addresses several barriers described in Section 2. The primary barrier is lack of IT skills. Instead of asking older adults to participate in a preceding ICT course, which will make the game less spontaneous, we opted for the passive assistance of a junior player who is teamed with a senior player. Studies conducted by Ng (2017) shown that peer tutor model is at least as efficient as formal classes. The presence of assistants was also a part of social support, which is a crucial factor for older adults to learn technology (Woodward et al., 2013).

We opted for matching older adults with young students, because according to the contact theory (Pettigrew, 1998) carefully crafted interaction between groups helps to overcome stereotypes and prejudices. Young people with proficiency in using ICT are also more easily available, which makes the design more realistic for practical application on a larger scale.

To overcome the lack of motivation, an invitation was designed as a short and intriguing movie that presented a crime story[3] which needs to be solved by participants during the game. Other barriers, such as lack of opportunity and lack of companions, are also addressed, as the game creates opportunities and motivates seniors to participate socially (by pairing seniors with junior players) as well as learning and physical activity.

---
[3] https://youtu.be/nclX8Y3lcVE

The main barrier that is not fully addressed by the design of the location-based game is lack of physical mobility. The design of the game is based on an itinerary - in our use-case, it was a distance of approximately 2000 meters in total. This distance can of course be reduced, and the game can be played in various locations (for example, indoors in historical landmark buildings). However, in our use-case, one senior decided that he was not able to walk the itinerary planned by the game. Instead, this senior used a car to drive short distances from one game-station to another, taking his junior partner as a passenger.

*Learning ICT skills by doing.*
The approach of the game is to lower the barrier to start using the game, offering the assistance of a young volunteer, while teaching ICT in the process. The purpose of the tool is to help older adults overcome this barrier beyond the boundaries of the game, providing them with the tool to use other IT services. "Scroll Around Yesterday" was designed to teach older adults the following skills:

- Navigating on a map using GPS location
- Connecting to a Wi-Fi hotspot
- Skanning a QR code
- Taking a panoramic picture using a tablet
- Searching for information on the Web using a tablet
- Playing puzzle games on a tablet (touch-based interface)

These skills were taught to seniors by asking them to perform in-game tasks (a different task at each game station) that required using these skills. Junior players provided passive assistance or advice (touching the tablet by a junior player was against the game rules).

*Supporting in-person participation.*
The location-based game seems to be the most useful for independently-living older adults. Instead of compensating for mobility constraints, it helps older adults to keep physically active and contributes to preventing mobility issues. The game by itself is a good motivator for taking a medium-length walk. The game story can be adapted so that the game can be used in any location. In-game tasks may also be changed from tasks that aim at teaching ICT techniques to tasks that require seniors to do physical exercises (this would not affect the overall game design. In this case, junior players can be health professionals who can assist seniors in physical exercises.)

*Creating opportunities for intergenerational interactions.*
The location-based game was successful not only in lowering technical barriers for seniors, but also in encouraging interpersonal and intergenerational interactions. The game was played simultaneously by several senior-junior pairs (in the case of our use-case, 15 pairs for the first edition and 12 pairs for second edition) in the same physical locations (various players follow the same itinerary, although they may take various amounts of time, or even use various means of communication).

### 1.4.3 Studies and Findings

The gameplay of Stroll Around Yesterday described as our use-case was held twice in Warsaw in the area of the Constitution Square (Southern part of the city center) on October 4th 2015 and May 14th 2016. An average older adult player was 69 years old (the oldest player was 86 years old) and 20 of 27 older adults' participants were woman.

*Effectiveness of Teaching ICT Skills to Seniors.*
To evaluate the effectiveness of the learning approach used in the game, we have asked both groups (seniors and juniors) to choose the most accurate description of what happened on each stage of the game on a 5-point scale from "junior completed the task alone" to "senior completed the task without any assistance". The results were rather consistent and very encouraging. In very few cases the evaluations by two parties differed by more than one category. In most cases the senior completed the tasks instructed by her/his partner. The most problematic task appeared to be establishing and verifying the connection to the Wi-Fi, as many seniors asked their partners for direct assistance. We should also add that the general opinion from the above-mentioned demonstrative game edition proved that tasks were not so easy to perform. Additionally, an interesting conclusion regarding learning and self-awareness is that seniors tend to underestimate their performance.

*Effectiveness of Improving Intergenerational Perception.*
The effectiveness of using "Stroll Around Yesterday" for fostering positive changes in the mutual perception of different generations was evaluated using a short survey that asked seniors and juniors to evaluate a "general other" (an average person that belongs to the specific group without pointing to particular person) from the other age group (seniors evaluated juniors, and vice-versa). The evaluation was done in several dimensions suggested by contact theory (Allport, 1954; Pettigrew, 1998). In all dimensions (i.e., passive vs. active, suspicious vs. trustful, dependent vs. independent, uncooperative vs. cooperative, defensive vs. aggressive), the median perception of the general junior by seniors and of the general senior by juniors has improved (for some dimensions, by over two points on the Likert scale). Detailed results are presented in Kopec et al. (2017).

## 1.5 Lessons learned

From this research we derive a number of findings that are relevant to the overall research on technology for social participation in later life:

*Technology barriers can be overcome*
Even in relatively complex applications and for both "younger old" (60+) and "older old" adults (85+), technology can be adopted and accepted. We showed that following design guidelines and 'learning by doing' can be effective approaches. However, in this context we also learned that there is a tradeoff to be carefully managed between familiarity (adopting metaphors that are consistent with a person's experiences), accessibility and aesthetics. The latter aspect is sometimes neglected in favor of functionality, and this is indeed one of the reasons for failure of much of the technology intended for people as they age (Consel, 2018).

*Technology can be a favourable platform for social participation*
Technology for social participation can actually enable older adults to participate in social activities, either in co-located or virtual settings. We have also seen that technology can provide further measurable benefits in terms of participation in the presence of persuasion techniques – which in our use cases were standard techniques that have been known to be effective across various age groups.

*Social context can be a driver for participation*
Engaging in activities with others (even in virtual form) not only creates opportunities for new social interactions, but is also a key motivating ingredient for participation. For example, we have seen that older adults prefer participating in training sessions with others, and that by training in a social context they also engaged in more training sessions that those training individually.

*ICT can help build bridges between generations*
Carefully crafted interactions between older adults and students based on contact theory can not only boost learning process but also help overcome existing prejudices. Taken together, this means that technology can be very effective to both enable and motivate participation in later life. It is important to highlight, however, some limitations and especial considerations in designing and deploying social participation technology

*Technology alone is not a guarantee for social interactions*
We observed this especially in real time contextual interactions in virtual environments. Users did not perceive this type of interaction as useful and was ultimately the least used feature in the trials. In the study in Russia (Nikitina et al., 2018.), we also observed low levels of group interactions using public messages when participants featured low level of group *cohesion*, meaning when most did not know each other before the intervention. This points to the need to go beyond enabling social interactions to explore if and how technology can incorporate strategies to stimulate virtual as well as co-located interactions. Recent work in this direction is exploring how technology can foster friendship by leveraging on common life points (Ibarra et al., 2018a), and support reconnecting with old friends by facilitating incremental and informed interactions in a way that is safe for everybody and less socially-awkward (Ibarra et al. 2018b).

*Cultural differences in social interactions*
Related to the previous point, we observed different social interaction patterns when comparing the results from the trials with Italian (Baez et al., 2017) and Russian (Nikitina et al., 2018) older adults, under the same conditions. While Italians naturally engaged in community building and preferred exchanging public messages, Russian older adults engaged very little in community building and shared very few public messages, limiting their exchanges to private channels. This is an indication that technology designers should consider cultural differences in enabling and stimulating social interactions, as also shown in prior studies (Neves, Franz, Judges, Beermann, & Baecker, 2017).

*Extremely diverse target group*
In addition to cultural differences we also observed high differences in cognition, skills, health and fitness level. It makes the process of designing applications more demanding and, at the same time, limits potential benefits. Moreover, a simple yet robust and unobtrusive heuristics (based on a significant amount of sensitive data) are required for matching the right tool with people. Objective data about fitness level or health, even if available, might not solve the problem because older adults have developed many strategies to deal with their limitations (e.g., driving a car instead of walking during the location-based game).

*Managing technical problems and frustrations*
During the trials we experimented technical issues, especially due to Internet connection issues. This proved to be a frustrating experience for some participants, requiring us to provide a support line to address the issues. In our experience, a deployment test in real settings and real users prior to the trial can help anticipate potential issues and refine technical support procedures.

*Technology designed specifically for older adults may cause fear of stigmatization*
We observed that especially for "young" older adults (60+) there is a strong resistance to using applications that are labeled as "for older adults" – e.g., some participants of the location-based game seemed to be a bit disappointed that they did not use iPads during the game because iPads are what they grandchildren use.

*Scalability might be an issue*
ICT solutions typically scale well with the number of users in terms of costs. However, especially in the trial phase, this is rarely true in technologies promoting social participation in later life, where the cost and effort grows almost linearly with the number of participants due to setup, training, and management needs. Thus, despite the benefits in enabling and motivating social participation, both our studies as well as the literature still fall short in exploring the impact of sustained use of participation technology and its effect on isolation, participation, and ultimately on wellbeing. Future work should focus not only on feasibility but also on collecting evidence that can better characterize the effects and benefits of technology for larger groups of older people, as well as providing better guidance on how to address the challenges in designing, developing and deploying technology for diverse and possibly vulnerable populations (Baez & Casati, 2018).

**Acknowledgments**

This work has received funding from the EU Horizon 2020 research and innovation programme under the Marie Skodowska-Curie grant agreement No 690962.